\documentclass[twocolumn]{revtex4}

\usepackage{graphicx}

\begin{document}


\title{Electron transport in parallel quantum wires with random potentials}

\author{Yousuke Takeuchi}
\affiliation{%
Department of Quantum Matter, ADSM, Hiroshima University\\
Higashi-Hiroshima 739-8530, Japan
}%

\author{Hiroyuki Mori}
\affiliation{
Graduate School of Engineering, Tokyo Metropolitan Institute of Technology\\
Tokyo 191-0065, Japan
}%

\date{\today}

\begin{abstract}
We study an electron transport property in two parallel quantum wires with random potentials.  Assuming the same microscopic parameters for both wires, we focus on the relationship between inter-wire interaction and electron backward scattering by random potentials at low energy regime. Our analytical and numerical calculations show that the Drude weight, a measure of the electron transport, is influenced by inter-wire interaction and random potential independently, and little coupling between those two is observed, which is in contrast to a deep relationship between up- and down-spin interactions and random potentials in a single wire. It leads to that inter-wire interactions do not have a great influence on the Anderson localization in each wire.
\end{abstract}

\maketitle
The problem of electronic transport has been one of the major issues in the history of condensed matter physics. There are always two key factors: electron-electron interaction and random potential. The importance of their roles on the electronic transport is extremely high in low dimensions because of large quantum fluctuations and strong localizations.

In this paper we pick up a one-dimensional system of parallel quantum wires. If no hopping between the wires is allowed, electrons on one wire recognize the presence of the neighboring wire only through inter-wire coulomb interaction.

In one-dimensional system, interacting electrons form so-called Luttinger liquid and the important theoretical works have been done on its transport properties. Kane and Fisher \cite{prl68} and Furusaki and Nagaosa \cite{prb47} studied Luttinger liquid placed in a single (and double) barrier system and showed the various interesting aspects of conductance.

Luttinger liquid with random potential was studied, for example in Ref.[3-5]. Electrons interacting each other with repulsive force are shown all localized even by an infinitesimally weak random potential \cite{prb37}. There is in fact an interplay between the interaction and the random potential, but the transport property would not reflect it because the system is always an insulator. The interplay becomes significant in considering the transport in finite size systems that could be metal \cite{prb51}. Reference \cite{prb51} claims that repulsive interaction would suppress the effects of random potentials and hence enhance the electronic current.

If we put two equivalent wires in a parallel position, another interesting phenomena is expected. When no electron transferring between the wires is assumed, the electrons on each wire are confined in the line, but the inter-wire interactions couple the electrons of both wires. Electronic current in one wire would drag the electrons in the other wire, which is called coulomb drag effect. For the parallel wires, some theoretical works have been done concerning the temperature and electron reservoir dependence of transresistance, transconductance and transsusceptibility for clean wires as well as wires with periodic potentials [6-8].

In this paper we focus our attentions on, if any, the interplay between inter-wire interaction and random potential. We would like to see how the electronic transport in a dirty wire is modified by the neighboring wire that is also dirty. For this purpose we calculated Drude weight of a wire in the presence of neighboring wire and try to seen how it varies by changing both the interaction and random potential, expecting it to provide us informations on whether the inter-wire interaction has any effect on the Anderson localization within the wire.

The model we consider here describes spinless electrons in two identical parallel one-dimensional wires with random potentials. It could be interpreted that we are treating parallel wires in a strong magnetic field that polarizes all the spins. Inter-wire and intra-wire interactions between electrons are also assumed. The Hamiltonian is then given by
\begin{eqnarray}
{\cal H}&=&-t\sum_{i,w}(c_{i,w}^\dagger c_{i+1,w}+h.c.)+U\sum_{i}n_{i,+}n_{i,-}\nonumber\\
& &{}+V\sum_{i,w}n_{i,w}n_{i+1,w}+\sum_{i,w}R_w(i)n_{i,w},\label{1}
\end{eqnarray}
where $t$ is the hopping integral, $U$ is the inter-wire interaction, $V$ is the intra-wire interaction, $R_w$ is random potential on the site $i$ and $w=\pm$ presents the wire index. 

In the low energy regime, the kinetic-energy part becomes 
\begin{equation}
{\cal H}_0=\pi{\mathit v}_FL^{-1}\sum_{r,w,k}\rho_{r,w}(k)\rho_{r,w}(-k),\label{2}
\end{equation}
where $r=\pm$ is an index for the right/left moving electrons, ${\mathit v}_F=2t \sin(k_F)$ is the Fermi velocity, $\rho_{r,w}(k)$ is the electron density in the momentum space. The small momentum transfer part of intra- and inter-wire interactions is given by 
\begin{eqnarray}
{\cal H}_f\!\!\!&=&\!\!\!L^{-1}\sum_{k,w,w'}(g_2\delta_{w,w'}+\bar{g}_2\delta_{w,-w'})\rho_{+,w}(k)\rho_{-,w'}(-k)\nonumber \\
\!\!\!&+&\!\!\!(2L)^{-1}\!\!\sum_{r,k,w,w'}\!(g_4\delta_{w,w'}+\bar{g}_4\delta_{w,-w'})\rho_{r,w}(k)\rho_{r,w'}(-k),\nonumber\\
\label{3}
\end{eqnarray}
where $g_2=2V(1-\cos(2k_F))$ and $g_4=2V$ are the coupling constants of intra-wire interactions, and $\bar{g}_2=U$ and $\bar{g}_4=U/2$ are of inter-wire interaction. The $2k_F$ backward scattering between the wires is given by
\begin{equation}
{\cal H}_b=\bar{g}_1\sum_w\int dx \psi_{+,w}^\dagger(x)\psi_{-,-w}^\dagger(x)\psi_{+,-w}(x)\psi_{-,w}(x),\label{4}
\end{equation}
where $\bar{g}_1=U$. The total Hamiltonian of clean wires is then given by ${\cal H}_c={\cal H}_0+{\cal H}_f+{\cal H}_b$, which is obviously equivalent to an isospin-1/2 single-channel system. 

Following the standard procedure of bozonization, we represent fermionic fields $\psi_{r,w}(x)$ as $\psi_{r,w}(x)=(2\pi \alpha)^{-1/2}\exp[ir\{k_Fx-\phi_w(x)\}+i\theta_w(x)]$ by introducing bosonic fields $\phi_w(x)=-i\pi L^{-1}\sum_q \frac{e^{-iqx-\alpha|q|/2}}{q}[\rho_{+,w}(q)+\rho_{-,w}(q)] $ and their conjugates $\Pi_w(x)=L^{-1}\sum_q e^{-iqx-\alpha|q|/2}[\rho_{+,w}(q)-\rho_{-,w}(q)]$, where $\alpha=+0$ and $\partial_x \theta_w(x)=\pi \Pi_w(x)$. In this bozon representation and the transformations $\phi_{c\pm}=(\phi_+\pm\phi_-)/\sqrt{2},\Pi_{c\pm}=(\Pi_+\pm\Pi_-)/\sqrt{2}$, the Hamiltonian becomes
\begin{eqnarray}
{\cal H}_c&=&\frac{u_{c+}}{2\pi}\int dx [K_{c+}\pi^2\Pi^2_{c+}+\frac{1}{K_{c+}}(\partial_x\phi_{c+})^2] \nonumber\\
&&{}+\frac{u_{c-}}{2\pi}\int dx [K_{c-}\pi^2\Pi_{c-}^2+\frac{1}{K_{c-}}(\partial_x \phi_{c-})^2]\nonumber \\
&&{}+\frac{2\bar{g}_1}{(2\pi\alpha)^2}\int dx \cos(\sqrt{8}\phi_{c-}),\label{5}
\end{eqnarray}
where
\begin{eqnarray}
u_{c\pm}&=&\mathit{v}_F[(1+\frac{g_4\pm\bar{g}_4}{2\pi \mathit{v}_F})^2-(\frac{g_2\pm\bar{g}_2}{2\pi\mathit{v}_F})^2]^{1/2},\label{6}\\
K_{c\pm}&=&[\frac{2\pi\mathit{v}_F+g_4\pm\bar{g}_4-g_2\mp\bar{g}_2}{2\pi\mathit{v}_F+g_4\pm\bar{g}_4+g_2\pm\bar{g}_2}]^{1/2}.\label{7}
\end{eqnarray}
When the interactions are weak, $u_{c\pm}$ and $K_{c\pm}$ are given by
\begin{eqnarray}
u_{c\pm}&\sim&v_F(1+\frac{V}{\pi v_F}\pm \frac{U}{4\pi v_F}), \label{ucpm}\\
K_{c\pm}&\sim&1-\frac{V}{\pi v_F}(1-\cos(2k_F))\mp\frac{U}{2\pi v_F}. \label{Kcpm}
\end{eqnarray}
We now introduce quenched random potentials, which are assumed to be parameterized by two Gaussian random fields $\eta_w$ and $\xi_w$, where $\eta$ is real and $\xi$ is  complex. These two fields represent respectively the forward and backward electron scattterings from the random potentials, and have the Gaussian distributions:
\begin{eqnarray}
P_{\eta_w}&=&\exp[-(2\Delta_f)^{-1}\int dx \eta_w^2(x)] \label{8a}\\
P_{\xi_w}&=&\exp[-(2\Delta_b)^{-1}\int dx |\xi_w(x)|^2].\label{8}
\end{eqnarray}
The scattering terms with the random potentials are given by 
\begin{eqnarray}
{\cal H}_{\mathrm{frand}}&=&\sum_w \int dx \eta_w(x)[\rho_{+,w}(x)+\rho_{-,w}(x)],\label{9}\\
{\cal H}_{\mathrm{brand}}&=&\sum_w \int dx[\xi_w(x)\psi_{+,w}^\dagger(x)\psi_{-,w}(x)\nonumber\\
& &{}\hspace{1.3cm}+\xi_w^*(x)\psi_{-,w}^\dagger(x)\psi_{+,w}(x)].\label{10}
\end{eqnarray}
It should be noted that introducing random potentials on each wire makes the present model variant from its resemblance, the isospin-1/2 single-channel model. This is because in the former case the profiles of the potentials that the electrons feel on each wire are different even though their statistical characteristics (average depth and standard deviation) are the same. In the latter model, on the other hand, the electrons of up and down spins feel exactly the same potentials. Our two-wire model with random potentials actually corresponds to an isospin-1/2 single-channel model with random potentials \textit{and} a random magnetic field.

Using the conventional replica trick, we integrate out the random field and obtain the following replica actions:
\begin{eqnarray}
\overline{S}&=&\overline{S_0}+\overline{S_1}+\overline{S_2},\label{11}\\
\overline{S_0}&=&\sum_i^n \int d\tau dx \frac{1}{2\pi}\nonumber\\
& &{}\!\!\!\times(\frac{1}{K_{c+}}[\frac{1}{u_{c+}}(\partial_\tau \tilde{\phi}_{c+}^i(x,\tau))^2+u_{c+}(\partial_x \tilde{\phi}_{c+}^i(x,\tau))^2]\nonumber \\
& &{}\!\!\!+\frac{1}{K_{c-}}[\frac{1}{u_{c-}}(\partial_\tau \tilde{\phi}_{c-}^i(x,\tau))^2+u_{c-}(\partial_x \tilde{\phi}_{c-}^i(x,\tau))^2]),\nonumber\\
\label{12}\\
\overline{S_1}&=&-\sum_{i,j}^n\int d\tau d\tau' dx dx' \frac{{\bar g}_1^2}{16\pi^4\alpha^2}\nonumber\\
& &{}\times\cos(\sqrt{8}[\tilde{\phi}_{c-}^i(x,\tau)-\tilde{\phi}_{c-}^j(x',\tau')])\nonumber\\
& &{}\times e^{-4\Delta_f(\frac{K_{c-}}{u_{c-}})^2|x-x'|},\label{13}\\
\overline{S_2}&=&-\sum_{i,j}^n\int d\tau d\tau' dx \frac{\Delta_b}{\pi^2\alpha^2}\nonumber\\
& &{}\times\cos(\sqrt{2}[\tilde{\phi}_{c-}^i(x,\tau)-\tilde{\phi}_{c-}^j(x,\tau')])\nonumber\\
& &{}\times\cos(\sqrt{2}[\tilde{\phi}_{c+}^i(x,\tau)-\tilde{\phi}_{c+}^j(x,\tau')]).\label{14}
\end{eqnarray}
where $i$ and $j$ are the replica indices. To derive the replica actions, we have shifted the bozon field as 
\begin{eqnarray}
\tilde{\phi}_{c+}&=&\phi_{c+}+\tilde{\eta}_{\mathrm{tot}}(x),\\\nonumber
\tilde{\eta}_{\mathrm{tot}}(x)&=&-\frac{K_{c+}}{\sqrt{2}u_{c+}}\int^x dz\eta_{\mathrm{tot}}(z), \\\nonumber
\tilde{\eta}_{\mathrm{rel}}(x)&=&-\frac{K_{c-}}{\sqrt{2}u_{c-}}\int^x dz\eta_{\mathrm{rel}}(z),\\\nonumber
\eta_{\mathrm{tot}}(x)&=&\eta_{+}(x)+\eta_{-}(x),\\\nonumber
\eta_{\mathrm{rel}}(x)&=&\eta_{+}(x)-\eta_{-}(x).\nonumber
\end{eqnarray}
For this action, we derived the following renormalization equations:
\begin{eqnarray}
\frac{dK_{c+}(l)}{dl}&=&-\mu \frac{K_{c+}^2}{u_{c+}^2}(K_{c-}+K_{c+})\Delta_b(l),\label{15}\\
\frac{du_{c+}(l)}{dl}&=&-\mu\frac{1}{u_{c+}}K_{c+}(K_{c-}+K_{c+})\Delta_b(l),\label{16}\\
\frac{dK_{c-}(l)}{dl}&=&-\mu\frac{K_{c-}^2}{u_{c-}^2}[(K_{c-}\!+\!K_{c+})\Delta_b(l)\!+\!\frac{2a}{\pi^2}K_{c-}\bar{g}_1^2(l)],\label{17}\nonumber\\\\
\frac{du_{c-}(l)}{dl}&=&-\mu \frac{1}{u_{c-}}K_{c-}(K_{c-}+K_{c+})\Delta_b(l),\label{18}\\
\frac{d\Delta_b(l)}{dl}&=&(3-K_{c-}-K_{c+})\Delta_b(l),\label{19}\\
\frac{d\bar{g}_1(l)}{dl}&=&2(1-K_{c-})\bar{g}_1(l),\label{20}
\end{eqnarray}
where $\mu=a/3\pi$. 

We hereafter focus on the Drude weight $D_w$ of wire $w(=\pm )$, which describes the electronic transport within the wire $+(-)$ in the presence of the neighboring wire $-(+)$. The Drude weight $D_{\pm}$ can be determined by the current-current correlation function $\langle J_{\pm}^2\rangle=\langle (J_{c+}\pm J_{c-})^2/2\rangle$, where $J_{c\pm}$ is symmetric/antisymmetric current densities. Since we assumed qualitative equivalence of both wires, we have $D_+=D_-\equiv D$ and therefore
\begin{equation}
D=(\langle J^2_{c+}\rangle +\langle J^2_{c-}\rangle )/2.
\end{equation}
With no inter-wire interaction and random potential, we get from the quadratic form of the current $\partial_{\tau}\phi_{c\pm}$ in the action
\begin{equation}
D=K_{c+}u_{c+}+ K_{c-}u_{c-},
\end{equation}
omitting the irrelevant numerical factors. The qualitative change of $D$ by the inclusion of the interaction and the random potential can be seen from the behaviors of $K$ and $u$ in the RG equations.

At first, $g_{1\perp}$ always makes $K_c$ smaller as the renormalization proceeds in Eq.(\ref{17}), indicating that the presence of the inter-wire interaction gives the smaller Drude weight irrespective to the sign of the interaction. This is the dragging effect, which is stronger for larger interaction.

Although the full dependence of Drude weight on the microscopic parameters has to be calculated by numerically integrating the RG equations, we take a rough view on the behavior of the parameter $\Delta_b$ neglecting the $l$ dependence of the $K$ on the r.h.s of Eq. (\ref{19}). Integrating the equation from $l=0$ to $l=\ln (L/a)$ we get
\begin{equation}
\Delta_b =\Delta_{b0}(L/a)^{1+2V[1-\cos(2k_F)]/\pi \mathit{v}_F},\label{22}
\end{equation}
where we used Eq. (\ref{Kcpm}) and $\Delta_{b0}$ is the initial value. Note that the inter-wire interaction $U$ has no direct effect on the random potential parameter $\Delta_b$, which means that the major role of $U$ on the electronic transport is to drag the neighboring wire electrons.

This is quite different from the case of the spin-full single wire model which is equivalent to the present model until the random potentials are introduced as stated above. The interaction dependence of the random potential parameter for the spin-full single wire is given by the following form (see Ref.\cite{prb51}). 
\begin{equation}
\Delta=\Delta_{0}(L/a)^{1-U/(\pi \mathit{v}_F)+2V[1-2\cos(2k_F)]/\pi \mathit{v}_F},\label{23}
\end{equation}
where $U$ and $V$ are the on-site and nearest-neighbor interactions, respectively. The important difference between Eq. (\ref{22}) and Eq. (\ref{23}) is the appearance of U in the latter. In the spin-full single wire, $U$, the interaction between up and down spins, would effectively increase the random potential parameter $\Delta$, which can be understood by the following argument: The spin-1/2 electron system in one dimension has the SDW-dominated ground state and the SDW fluctuation becomes larger with increasing $U$, whereas it suppresses CDW fluctuations. Therefore the pinning effect due to impurities is weaker for larger $U$. Since the last term in the r.h.s. of Eq. (\ref{17}) does not exist in the RG equations for the spin-full single wire case, meaning the absence of the dragging effect, the influence of $U$ on the Drude weight $D$ appears solely through $\Delta$. Consequently $U$ would suppress $\Delta$ and therefore $D$. In other words the repulsive interactions between up- and down- spins help the transport. The different roles of interactions on the localization in our model and the spinfull single-wire model can also be seen in the localization length $\xi_{\mathrm{loc}}$. Following Ref. \cite{prb56-58}, we can estimate $\xi_{\mathrm{loc}}$ as follows,
\begin{eqnarray}
\xi_{\mathrm{loc}}/a \sim
\left\{
\begin{array}{l}
  (\xi^{(0)}_{\mathrm{loc}}/a)^{1-2V[1-\cos(2k_F)]/\pi \mathit{v}_F}\\
  \hspace{1.5cm}(\mbox{spinless two-wire model})\\\\
  (\xi^{(0)}_{\mathrm{loc}}/a)^{1+U/(\pi \mathit{v}_F)-2V[1-\cos(2k_F)]/\pi \mathit{v}_F}\\
  \hspace{1.5cm}(\mbox{spinfull single wire model})\\
\end{array}
\right. 
\end{eqnarray}
where $a$ is the lattice constant and $\xi^{(0)}_{\mathrm{loc}}$ is the localization length of non-interacting system.

In order to check the above result numerically, we performed Monte Carlo calculations for the Hamiltonian (\ref{1}) using the world-line algorithm. The random potential on each site is assumed to be a random number uniformly generated between $-R$ and $R$. The simulation was performed in the zero winding number mode, based on the periodic boundary condition. While it is necessary to calculate the current-current correlation function to obtain Drude weight, its zero-frequency part $<  J(\omega )J(-\omega )> $ is always zero because we worked in the zero winding number space. We therefore extrapolated $<  J(\omega )J(-\omega )> $ to $\omega\rightarrow 0$ limit and determined the Drude weight \cite{prl65}. The simulations are performed for wires of 20 sites and 8 particles each. The random potential average were took over 100 samples to achieve satisfactory precision.

Figure 1 shows the Drude weight as a function of the inter-wire interaction $U$, at the temperature $(2/9)t$. The intra-wire interaction $V$ is fixed: $V=(2/3)t$. We see the monotonic reduction of $D$ for either increasing $U$ or increasing $R$. Let us normalize the Drude weight with the value of the clean system, {\it i.e.} $D(U,R)/D(U,R=0)$. Then we find the normalized $D$ has little dependence on $U$ (see the inset of Fig.1), which supports an idea that $D(U,R)$ can be approximately written as a product of two functions, one of $U$ and the other of $R$. If we write this as $D(U,R)\sim F_1(U)F_2(R)$, $F_1$ and $F_2$ are both decreasing functions. The point here is that $U$ and $R$ are not coupled in the Drude weight, which is exactly what we have on the RG calculations.

In conclusion we studied a transport property of spin-polarized electrons moving in parallel one-dimensional wires with random potentials. We in particular focused on the Drude weight, a measure of the transport in one wire under the influence of the other. Analytical and numerical calculations were done to see how the Drude weight varies as a function of the inter-wire interaction or of the random potential. We found both the interaction and the potential reduce the Drude weight: the former is due to the coulomb drag and the latter to the localization. An interesting point is that there is no noticeable interplay between these two. This is in contrast to the case of full-spin electrons in a single wire, where interactions between up- and down-spin electrons would weaken the random potential effectively. Little effect of the inter-wire interaction on the strength of the random potential means that the property of Anderson localization in each wire has little dependence on the inter-wire interaction, although it is greatly affected by the intra-wire interaction and by the inter-wire hopping \cite{prb56-58}.

Acknowledgement

The authors would like to thank prof. T. Jo and prof. T. Oguchi for helpful discussions.
\bibliography{apssamp}

\begin{thebibliography}{}\label{sec:TeXbooks}
\bibitem{prl68}C. L. Kane and M. P. A. Fisher, Phys. Rev. Lett. \textbf{68}, 1220 (1992)

\bibitem{prb47}A. Furusaki and N. Nagaosa, Phys. Rev. B \textbf{47}, 3827 (1993); Phys. Rev. B \textbf{47}, 4631 (1993) 

\bibitem{prb37}T. Giamarchi and H. J. Schulz, Phys. Rev. B \textbf{37}, 325 (1988)

\bibitem{prb40}M. P. A. Fisher, P. B. Weichman, G. Grinstein and D. S. Fisher, Phys. Rev. B \textbf{40}, 546 (1989)

\bibitem{prb51}T. Giamarchi and B. S. Shastry, Phys. Rev. B \textbf{51}, 10915 (1995)

\bibitem{prb62}R. Klesse and A. Stern, Phys. Rev. B \textbf{62}, 16912 (2000)

\bibitem{prl81}Y. V. Nazarov and D. V. Averin, Phys. Rev. Lett. \textbf{81}, 653 (1998)

\bibitem{prl85}V. V. Ponomarenko and D. V. Averin, Phys. Rev. Lett. \textbf{85}, 4928

\bibitem{prb56-58}E. Orignac and T. Giamarchi, Phys. Rev. B. \textbf{56}, 7167 (1997) ; H. Mori, \textit{ibid} \textbf{58}, 12699 (1998)

\bibitem{prl65}G. G. Batrouni, R. T. Scalettar and G. T. Zimanyi, Phys. Rev. Lett. \textbf{65}, 1765
\end{thebibliography}

\begin{figure}
\includegraphics[width=6cm,height=5.4cm]{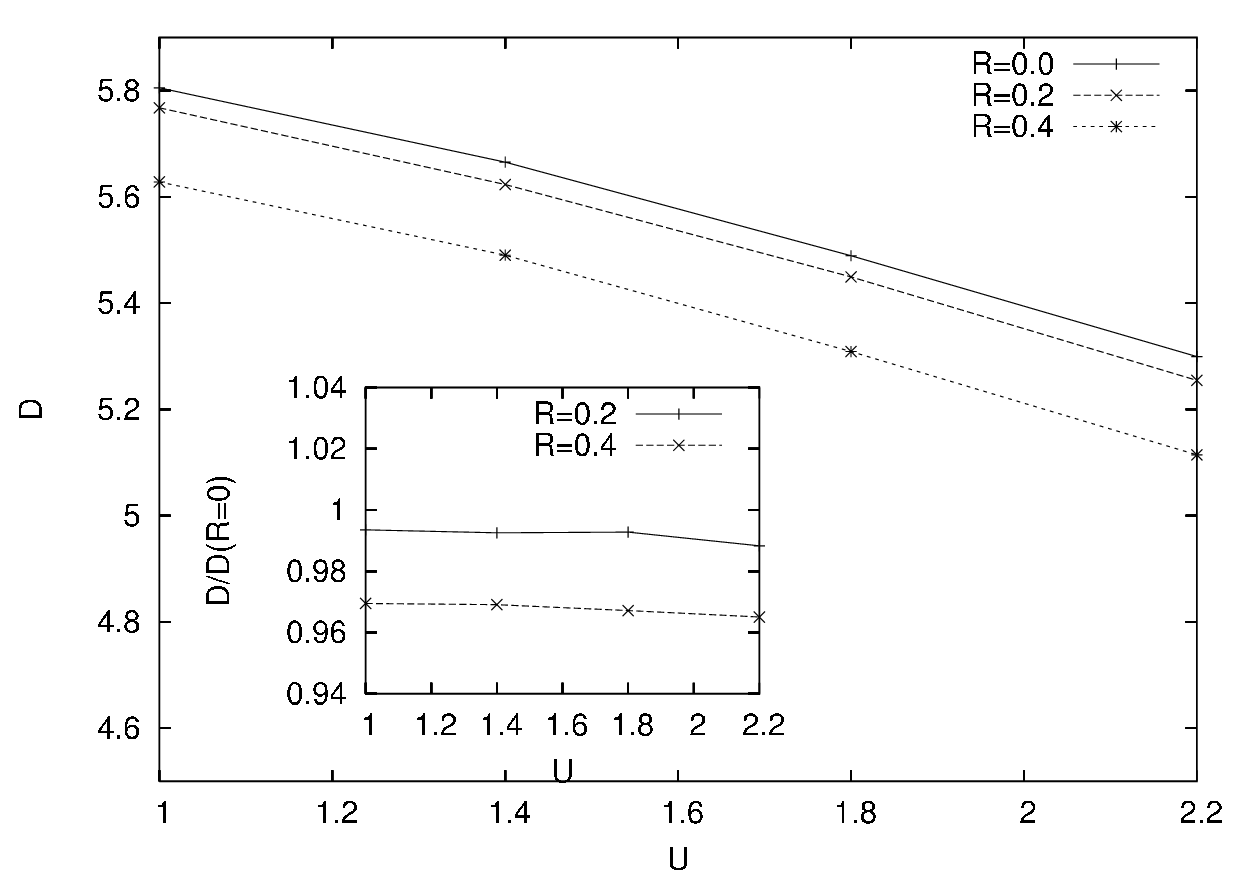}
\caption{\label{fig:epsart} Drude weight $D$ as a function of inter-wire interaction $U$ for the spinless double wire system. The random potential varies between $-R$ and $R$ and $t$ and $V$ are fixed to be 1.5 and 1 respectively.}
\end{figure}

\end{document}